\documentstyle[12pt,epsf]{article}

\title{%
\hfill
 \parbox{5cm}{\normalsize ICRR-Report-393-97-16\\
 \normalsize UT-779\\
 hep-ph/9707203}\\
\vspace{5ex}
  $CP$- and $T$-Violation Effects in Long Baseline Neutrino
  Oscillation Experiments%
\vspace{5ex}
}

\author{
 Masafumi Koike%
\thanks{e-mail address: {\tt koike@icrr.u-tokyo.ac.jp}}\\%
{\footnotesize \it%
 Institute for Cosmic Ray Research, University of Tokyo,
 Midori-cho, Tanashi, Tokyo 188, Japan%
}
\\
and
\\
Joe Sato%
\thanks{e-mail address: {\tt joe@icrr.u-tokyo.ac.jp}}\\%
 {\footnotesize \it%
   Department of Physics, University of Tokyo, Hongo, Bunkyo-ku, Tokyo
   133, Japan%
 }%
}

\date{}

\begin{document}
\maketitle

\begin{abstract}
  We examine how large $CP$- and $T$-violation effects are allowed in
  long baseline neutrino experiments with three generations of
  neutrinos, considering both the solar neutrino deficit and the
  atmospheric neutrino anomaly.  We considerd two cases: When we
  attribute only the atmospheric neutrino anomaly to neutrino
  oscillation and assume the constant transition probability of
  $\nu_{\rm e}$ to explain the solar neutrino deficit, we may have
  large $CP$-violation effect.  When we attribute both the atmospheric
  neutrino anomaly and the solar neutrino deficit to neutrino
  oscillation, we can see sizable $T$- violation effects.  In this
  case, however, we cannot ignore the matter effect and we will not
  see the pure $CP$- violation effect.  We also show simple methods
  how to separate pure $CP$ violating effect from the matter effect.
  We give compact formulae for neutrino oscillation probabilities
  assuming one of the three neutrino masses (presumably $\nu_{\tau}$
  mass) to be much larger than the other masses and the effective mass
  due to matter effect.  Two methods are shown: One is to observe
  envelopes of the curves of oscillation probabilities as functions of
  neutrino energy; a merit of this method is that only a single
  detector is enough to determine the presence of $CP$ violation.  The
  other is to compare experiments with at least two different baseline
  lengths; this has a merit that it needs only narrow energy range of
  oscillation data.
\end{abstract}

\section{Introduction}
\quad\,
The $CP$ and $T$ violation is a fundamental and important problem of
the particle physics and cosmology.  The $CP$ violation has been
observed only in the hadron sector, and it is very hard for us to
understand where the $CP$ violation originates from.  If we observe
$CP$ violation in the lepton sector through the neutrino oscillation
experiments, we will be given an invaluable key to study the origin of
$CP$ violation and to go beyond the Standard Model.

The neutrino oscillation search is a powerful experiment which can
examine masses and/or mixing angles of the neutrinos.  The several
underground experiments, in fact, have shown lack of the solar
neutrinos\cite{Ga1,Ga2,Kam,Cl} and anomaly in the atmospheric
neutrinos\cite{AtmKam,IMB,SOUDAN2}%
\footnote{%
  Some experiments have not observed the atmospheric neutrino
  anomaly\cite{NUSEX,Frejus}.%
}, strongly indicating the neutrino oscillation\cite{Fogli1, FLM,
  Yasuda}.  The solar neutrino deficit implies $10^{-5} \sim 10^{-4}
\;{\rm eV^2}$ as a difference of the masses squared ($\delta m^2$),
while the atmospheric neutrino anomaly suggests $\delta m^2$ around
$10^{-3} \sim 10^{-2} {\rm eV^2}$\cite{Fogli1,FLM,Yasuda}.

The latter encourages us to make long baseline neutrino oscillation
experiments.  Recently such experiments are planned and will be
operated in the near future\cite{KEKKam,Ferm}. It is now desirable to
examine whether there is a chance to observe not only the neutrino
oscillation but also the $CP$ or $T$ violation by long baseline
experiments\cite{Tanimoto,ArafuneJoe,Joe,AKS,MN,BGG}.

In this paper we review our papers\cite{ArafuneJoe,Joe,AKS} to show
how large violation effects of $CP$ and $T$ we may see in long
baseline neutrino experiments with three generations of neutrinos,
considering both the solar neutrino deficit and the atmospheric
neutrino anomaly.

If we are to attribute both solar neutrino deficit and atmospheric
neutrino anomaly to neutrino oscillation, it is natural to consider
that one of $\delta m^2$'s is in the range ${\cal O}(10^{-5} \sim
10^{-4} \;{\rm eV^2})$ and the other is in ${\cal O}(10^{-3} \sim
10^{-2} \;{\rm eV^2})$.  Recently, however, Acker and
Pakvasa\cite{AckPak} has argued that it is possible to explain both
experiments by only one $\delta m^2$ scale around ${\cal O}(10^{-3}
\sim 10^{-2} \;{\rm eV^2})$.  In the former case we cannot ignore the
matter effect\cite{Krastev} and will not see pure $CP$ violating
effect; pure $CP$ violating effect must be separated from the matter
effect.  On the other hand, almost pure $CP$ violating effect can be
seen in the latter case.

In sec.\ref{section:review} we briefly review $CP$ and $T$ violation
in neutrino oscillation.  We then consider how large $T$ and $CP$
violating effects can be.  The case where both $\delta m^2$'s are
${\cal O} (10^{-2} \;{\rm eV^2})$ are considered in
sec.\ref{section:comparable}.  Sec.\ref{section:disparate} treats the
``disparate'' case, where the two $\delta m^2$'s are ${\cal O}
(10^{-2} \;{\rm eV^2})$ and ${\cal O} (10^{-4} \;{\rm eV^2})$.  In
sec.\ref{section:summary} we summarize our work and give discussions.


\section{Formulation of $CP$ and $T$ Violation in Neutrino Oscillation}
\label{section:review}
\quad\,
Let us briefly review $CP$ and $T$ violation in neutrino oscillation
\cite{FukugitaYanagida,BilenkyPetcov,Pakvasa} to clarify our notation.

We assume three generations of neutrinos which have mass eigenvalues
$m_{i} (i=1, 2, 3)$ and mixing matrix $U$ relating the flavor
eigenstates $\nu_{\alpha} (\alpha={\rm e}, \mu, \tau)$ and the mass
eigenstates in the vacuum $\nu\,'_{i} (i=1, 2, 3)$ as
\begin{equation}
  \nu_{\alpha} = U_{\alpha i} \nu\,'_{i}.
  \label{Udef}
\end{equation}
We parameterize $U$\cite{ChauKeung,KuoPnataleone,Toshev}
as

\begin{eqnarray}
& &
U
=
{\rm e}^{{\rm i} \psi \lambda_{7}} \Gamma {\rm e}^{{\rm i} 
\phi \lambda_{5}} {\rm e}^{{\rm i} \omega \lambda_{2}} \nonumber 
\\
&=&
\left(
\begin{array}{ccc}
  1 & 0 & 0  \\
  0 & c_{\psi} & s_{\psi} \\
  0 & -s_{\psi} & c_{\psi}
\end{array}
\right)
\left(
\begin{array}{ccc}
  1 & 0 & 0  \\
  0 & 1 & 0  \\
  0 & 0 & {\rm e}^{{\rm i} \delta}
\end{array}
\right)
\left(
\begin{array}{ccc}
  c_{\phi} & 0 &  s_{\phi} \\
  0 & 1 & 0  \\
  -s_{\phi} & 0 & c_{\phi}
\end{array}
\right)
\left(
\begin{array}{ccc}
  c_{\omega} & s_{\omega} & 0 \\
  -s_{\omega} & c_{\omega} & 0  \\
  0 & 0 & 1
\end{array}
\right)
\nonumber \\
&=&
\left(
\begin{array}{ccc}
   c_{\phi} c_{\omega} &
   c_{\phi} s_{\omega} &
   s_{\phi}
  \\
   -c_{\psi} s_{\omega}
   -s_{\psi} s_{\phi} c_{\omega} {\rm e}^{{\rm i} \delta} &
   c_{\psi} c_{\omega}
   -s_{\psi} s_{\phi} s_{\omega} {\rm e}^{{\rm i} \delta} &
   s_{\psi} c_{\phi} {\rm e}^{{\rm i} \delta}
  \\
   s_{\psi} s_{\omega}
   -c_{\psi} s_{\phi} c_{\omega} {\rm e}^{{\rm i} \delta} &
   -s_{\psi} c_{\omega}
   -c_{\psi} s_{\phi} s_{\omega} {\rm e}^{{\rm i} \delta} &
   c_{\psi} c_{\phi} {\rm e}^{{\rm i} \delta}
\end{array}
\right),
\label{UPar2}
\end{eqnarray}
where $c_{\psi} = \cos \psi, s_{\phi} = \sin \phi$, etc.

The evolution equation for the flavor eigenstate vector in the vacuum
is
\begin{eqnarray}
 {\rm i} \frac{{\rm d} \nu}{{\rm d} x}
&=&
 - U {\rm diag}(p_1, p_2, p_3) U^{\dagger} \nu \nonumber \\
&\simeq&
 \left\{
  - p_1 +
  \frac{1}{2 E}
  U {\rm diag} (0, \delta m^2_{21}, \delta m^2_{31})
  U^{\dagger}
 \right\} \nu,
\end{eqnarray}
where $p_i$'s are the momenta, $E$ is the energy and $\delta m^2_{ij}
= m^2_i - m^2_j$.  Neglecting the term $p_1$ which gives an irrelevant
overall phase, we have
\begin{equation}
 {\rm i} \frac{{\rm d} \nu}{{\rm d} x}
=
 \frac{1}{2 E}
 U {\rm diag} (0, \delta m^2_{21}, \delta m^2_{31})
 U^{\dagger}
 \nu.
 \label{VacEqnMotion}
\end{equation}

Similarly the evolution equation in matter is expressed as
\begin{equation}
 {\rm i} \frac{{\rm d} \nu}{{\rm d} x}
 = H \nu,
 \label{MatEqn}
\end{equation}
where
\begin{equation}
  H
  \equiv
  \frac{1}{2 E}
  \tilde U
  {\rm diag} (\tilde m^2_1, \tilde m^2_2, \tilde m^2_3)
  \tilde U^{\dagger},
 \label{Hdef}
\end{equation}
with a unitary mixing matrix $\tilde U$ and the effective mass squared
$\tilde m^{2}_{i}$'s $(i=1, 2, 3)$.  (Tilde indicates quantities in
the matter in the following).  The matrix $\tilde U$ and the masses
$\tilde m_{i}$'s are determined by\cite{Wolf,MS}
\begin{equation}
\tilde U
\left(
\begin{array}{ccc}
  \tilde m^2_1 & & \\
  & \tilde m^2_2 & \\
  & & \tilde m^2_3
\end{array}
\right)
\tilde U^{\dagger}
=
U
\left(
\begin{array}{ccc}
  0 & & \\
  & \delta m^2_{21} & \\
  & & \delta m^2_{31}
\end{array}
\right)
U^{\dagger}
+
\left(
\begin{array}{ccc}
  a & & \\
  & 0 & \\
  & & 0
\end{array}
\right).
\label{MassMatrixInMatter}
\end{equation}
Here
\begin{eqnarray}
 a &\equiv& 2 \sqrt{2} G_{\rm F} n_{\rm e} E \nonumber \\
   &=& 7.56 \times 10^{-5} {\rm eV^{2}}
       \frac{\rho}{\rm g\,cm^{-3}}
       \frac{E}{\rm GeV},
 \label{aDef}
\end{eqnarray}
where $n_{\rm e}$ is the electron density and $\rho$ is the matter
density.  The solution of eq.(\ref{MatEqn}) is then
\begin{equation}
 \nu (x) = S(x) \nu(0)
 \label{nu(x)}
\end{equation}
with
\begin{equation}
 S \equiv {\rm T\, e}^{ -{\rm i} \int_0^x {\rm d} s H (s) }
 \label{Sdef}
\end{equation}
(T being the symbol for time ordering), giving the oscillation
probability for $\nu_{\alpha} \rightarrow \nu_{\beta} (\alpha, \beta =
{\rm e}, \mu, \tau)$ at distance $L$ as
\begin{eqnarray}
 P(\nu_{\alpha} \rightarrow \nu_{\beta}; E, L)
&=&
 \left| S_{\beta \alpha} (L) \right|^2.
 \label{alpha2beta}
\end{eqnarray}
The oscillation probability for the antineutrinos $P(\bar\nu_{\alpha}
\rightarrow \bar\nu_{\beta})$ is obtained by replacing $a \rightarrow
-a$ and $U \rightarrow U^{\ast} ({\rm i.e.\,} \delta \rightarrow
-\delta)$ in eq.(\ref{alpha2beta}).
 
We assume in the following that the matter density is independent of space
and time for simplicity%
\footnote{%
In case matter density spatially varies, one can use an averaged value
$\left< a \right>$ in place of $a$\cite{Joe}.  We have presented an
example of this replacement for the case of KEK/Super-Kamiokande
experiment in \ref{appendix:derivation}.  See
Fig.\ref{nonuniformMatter}.%
}.  In this case we have
\begin{equation}
  S(x) = {\rm e}^{ -{\rm i} H x}
 \label{simpleS}
\end{equation}
and
\begin{eqnarray}
 P(\nu_\alpha \rightarrow \nu_\beta; E, L)
 &=&
 \left|
  \sum_{i,j} \tilde U_{\beta i}
  \left( {\rm e}^{
   -{\rm i}\frac{L}{2E}
    {\rm diag}(0, \delta \tilde m_{21}^2, \delta \tilde m_{31}^2)
    }
  \right)_{ij}
  \tilde U^{\ast}_{\alpha j} \right|^2
 \\
 &=&
  \sum_{i,j}
   \tilde U_{\beta i} \tilde U^{\ast}_{\beta j}
   \tilde U^{\ast}_{\alpha i} \tilde U_{\alpha j}
  {\rm e}^{ -{\rm i} \frac{\delta \tilde m^2_{ij} L}{2 E} },
  \label{SimpleP}
\end{eqnarray}
where $\delta \tilde m^2_{ij} \equiv \tilde m^2_i - \tilde m^2_j$.

The $T$ violation gives the difference between the transition
probability of $\nu_\alpha \rightarrow \nu_\beta$ and that of
$\nu_\beta \rightarrow \nu_\alpha$\cite{BWP}:
\begin{eqnarray}
 & &
 P(\nu_\alpha \rightarrow \nu_\beta; E, L) -
 P(\nu_\beta \rightarrow \nu_\alpha; E, L)
 \nonumber \\
 &=&
 -4 ({\rm Im} \; \tilde U_{\beta 1} \tilde U^{\ast}_{\beta 2}
     \tilde U^{\ast}_{\alpha 1} \tilde U_{\alpha 2})
 (\sin\Delta_{21}+\sin\Delta_{32}+\sin\Delta_{13})
 \label{eq:BWP}
 \\
 &\equiv&
 4 J f,
 \label{eq:jf}
\end{eqnarray}
where 
\begin{eqnarray}
 \Delta_{ij}
 &\equiv&
 \frac{\delta \tilde m_{ij}^2 L}{2 E}
 =
 2.54
 \frac{
  (\delta \tilde m_{ij}^2 / 10^{-2}{\rm eV}^2) (L/100{\rm km})
 }{ (E/{\rm GeV}) },
 \label{eq:Delta-numerical}
 \\
 J
 &\equiv&
 -{\rm Im} \;
 \tilde U_{\beta 1} \tilde U^{\ast}_{\beta 2}
 \tilde U^{\ast}_{\alpha 1} \tilde U_{\alpha 2},
 \label{eq:Jdef}
 \\
 f
 &\equiv&
 \sin\Delta_{21} + \sin\Delta_{32} + \sin\Delta_{13}
 \label{eq:fdef}
 \\
 &=&
 -4 \sin \frac{ \Delta_{21} }{2} \sin \frac{ \Delta_{32} }{2}
    \sin \frac{ \Delta_{13} }{2}.
 \label{eq:fv}
\end{eqnarray}

The unitarity of $U$ gives
\begin{eqnarray}
  J = \pm \sin \tilde\delta \cos^2\tilde\phi \sin \tilde\phi \cos
  \tilde\psi \sin \tilde\psi \cos \tilde\omega \sin \tilde\omega
 \label{eq:J}
\end{eqnarray}
with the sign $+$ ($-$) for $\alpha, \beta$ in cyclic (anti-cyclic)
order ($+$ for $(\alpha,\beta) =(e,\mu), (\mu,\tau)$ or $(\tau,e)$).
In the following we assume the cyclic order for ($\alpha$, $\beta$)
for simplicity.

There are bounds for $J$ and $f$.  $J$ satisfies
\begin{equation}
 |J| \le \frac{1}{ 6\sqrt{3} },
 \label{eq:Jmax}
\end{equation}
where the equality holds for $|\sin\tilde\omega| = 1/\sqrt{2},
|\sin\tilde\psi| = 1/\sqrt{2}, |\sin\tilde\phi| = 1/\sqrt{3}$
and $|\sin\tilde\delta|=1$.  On the other hand, $f$ satisfies\cite{Cabibbo}
\begin{eqnarray}
  |f| \le \frac{ 3 \sqrt3 }{2},
\label{eq:fmax}
\end{eqnarray}
where the equality holds for
$\Delta_{21} \equiv \Delta_{32} \equiv 2\pi/3 $ (mod $2\pi$).

In the vacuum the $CPT$ theorem gives the relation between
the transition probability of
anti-neutrino and that of neutrino,
%
\begin{eqnarray}
 P (\bar\nu_\alpha \rightarrow \bar\nu_\beta; E, L) =
 P (\nu_\beta \rightarrow \nu_\alpha; E, L),
\end{eqnarray}
which relates $CP$ violation to $T$ violation:
\begin{eqnarray}
&&
 P(\nu_\alpha \rightarrow \nu_\beta; E, L) -
 P(\bar\nu_\alpha \rightarrow \bar\nu_\beta; E, L)
 \nonumber \\
&=&
 P(\nu_\alpha \rightarrow \nu_\beta; E, L) -
 P(\nu_\beta \rightarrow \nu_\alpha; E, L).
\label{eq:cpt}
\end{eqnarray}

\section{Cases of Comparable Mass Differences}
\label{section:comparable}
\quad\,
Recently Acker and Pakvasa\cite{AckPak} argued the possibility of
explaining both solar neutrino experiments and atmospheric neutrino
experiments by one mass scale, $\delta m^2 \sim {\cal O}(10^{-3} \sim
10^{-2} \;{\rm eV^2}).$  We first examine this ``comparable mass
differences'' case.


We use a parameter set $(\delta m^2_{21},\delta m^2_{31}) = (3.8,1.4)
\times 10^{-2}$eV$^2$, $(\omega,\phi,\psi) =
(19^\circ,43^\circ,41^\circ)$ with arbitrary $\delta$, derived by
Yasuda\cite{Yasuda} through the analysis of the atmospheric neutrino
anomaly.  We need not distinguish $T$ violation and $CP$ violation in
this mass region since the matter effect is negligibly small.
Equation (\ref{eq:cpt}) is hence available.

Using eqs.(\ref{eq:jf}), (\ref{eq:J}) and (\ref{eq:cpt}), this
parameter set gives the $CP$-violation effect
\begin{eqnarray}
 P(\nu_\alpha \rightarrow \nu_\beta) -
 P(\bar\nu_\alpha \rightarrow \bar\nu_\beta)
=
 0.22 f(y) \sin\delta,
 \label{eq:CP1}
\end{eqnarray}
where
\begin{eqnarray}
  f(y) = ( \sin 3.8 y + \sin 2.4 y - \sin 1.4 y),
  \label{eq:CP1f}
\end{eqnarray}
and
\begin{eqnarray}
  y = 2.5 {(L/100{\rm km}) \over (E/{\rm GeV})}.
\end{eqnarray}

\begin{figure}
 \unitlength=1cm
 \begin{picture}(16,6)
  \unitlength=1mm
  \put(135,38){$y$}
  \put(0,75){$f(y) \equiv \sin 3.8 y + \sin 2.4 y - \sin 1.4 y$}
  \centerline{
   \epsfxsize=13cm
   \epsfbox{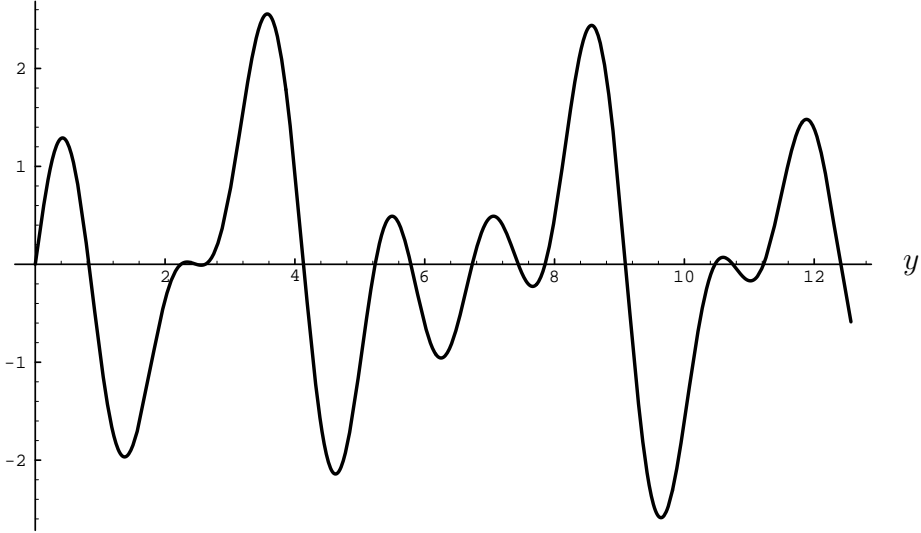}
  }
 \end{picture}
\caption{%
  Graph of $f(y)$ of eq.(\ref{eq:CP1f}).  There are high peaks
  (positive or negative) at $y = 0.42, 1.4, 3.6, 4.6 \cdots$.  Values
  of $f(y)$ at peaks averaged over energy spread of 10\% $\sim$ 20\%
  are $\left< f(0.42) \right> = 1.3 \sim 1.3, \left< f(1.4 ) \right>
  =-1.9 \sim -1.8, \left< f(3.6 ) \right> = 2.2 \sim 1.4, \left<
  f(4.6) \right> =-1.5 \sim -0.40\cdots$.}
\label{fig2}
\end{figure}

Figure \ref{fig2} shows the oscillatory part $f(y)$.  $f(y)$ has many
peaks showing the possibility to observe the large $CP$-violation
effect. For example, we may see very large difference between the
transition probabilities, $\left< P(\nu_\alpha \rightarrow \nu_\beta)
    - P(\bar\nu_\alpha \rightarrow \bar\nu_\beta) \right>_{20\%} \sim
0.4\sin\delta$ for $L = 250$ km (for KEK/Super-Kamiokande experiment)
and $E \sim$ 4.5 GeV corresponding to $y \sim 1.4$.  Hence it will be
possible to detect $CP$-violation effect if we have large
$\sin\delta$.

In general atmospheric neutrino anomaly indicates large mixing angles.
We may see a large $CP$-violation effect when we have
comparable mass differences.  In this respect we note that the long
baseline experiments are urgently desirable.

\section{Cases of Disparate Mass Differences}
\label{section:disparate}
\quad\,
Both solar neutrino deficit and atmospheric neutrino anomaly are
naturally explained by introducing two mass scales.  Solar neutrino
experiments suggest a mass difference of ${\cal O}(10^{-5} \sim
10^{-4} \;{\rm eV^2})$, while atmospheric neutrino measurements imply
${\cal O}(10^{-3} \sim 10^{-2} \;{\rm eV^2})$.  Here we consider this
``disparate mass difference'' case.  We see in this case that matter
effect given by eq.(\ref{aDef}) is the same order of magnitude as the
smaller mass scale.  Hence we cannot ignore matter effect.

\subsection{Transition Probabilities in Presence of Matter}
\label{subsection:trans-prob}
\quad\,
Let us derive simple expressions of oscillation probabilities
assuming $a, \delta m^2_{21} \ll \delta m^2_{31}$.
Decomposing $H = H_0 + H_1$ with
\begin{equation}
  H_0 = \frac{1}{2 E} U \left(
 \begin{array}{ccc}
  0 &   &  \\
    & 0 &  \\
    &   & \delta m^2_{31}
 \end{array}
 \right)
 U^{\dagger}
 \label{H0def} \\
\end{equation}
and
\begin{equation}
 H_1
=
 \frac{1}{2 E}
 \left\{
  U
  \left(
  \begin{array}{ccc}
   0 &                 &  \\
     & \delta m^2_{21} &  \\
     &                 & 0
  \end{array}
  \right)
  U^{\dagger} +
  \left(
  \begin{array}{ccc}
   a &   &  \\
     & 0 &  \\
     &   & 0
  \end{array}
  \right)
 \right\},
 \label{H1def}
\end{equation}
we treat $H_1$ as a perturbation and calculate eq.(\ref{simpleS})
up to the first order in $a$ and $\delta m^2_{21}$.  Defining
$\Omega (x)$ and $H_1 (x)$ as
\begin{equation}
 \Omega (x) = {\rm e}^{ {\rm i} H_0 x } S(x)
 \label{Omegadef}
\end{equation}
and
\begin{equation}
 H_1 (x) = {\rm e}^{ {\rm i} H_0 x} H_1 {\rm e}^{ -{\rm i} H_0 x},
 \label{H1(x)def}
\end{equation}
we have
\begin{equation}
 {\rm i} \frac{ {\rm d} \Omega}{ {\rm d} x }
=
 H_1 (x) \Omega (x)
 \label{Omegaeq}
\end{equation}
and
\begin{equation}
 \Omega (0) = 1,
 \label{OmegaInit}
\end{equation}
which give the solution
\begin{eqnarray}
 \Omega (x)
&=&
 {\rm T \, e}^{ -{\rm i} \int_{0}^{x} {\rm d} s H_1 (s)  }
 \nonumber \\
&\simeq&
 1 - {\rm i} \int_{0}^{x} {\rm d} s H_1 (s).
 \label{OmegaApprox}
\end{eqnarray}
We note the approximation (\ref{OmegaApprox}) requires
\begin{equation}
 \frac{a x}{2 E} \ll 1 \quad {\rm and} \quad
 \frac{\delta m^2_{21} x}{2 E} \ll 1
 \label{AppCond}.
\end{equation}
The equations (\ref{Omegadef}) and (\ref{OmegaApprox}) give\footnote{
We note the eq.(\ref{S0+S1}) is
correct for a case that the matter density depends on $x$.}

\begin{equation}
 S(x) \simeq {\rm e}^{ -{\rm i} H_0 x } +
             {\rm e}^{ -{\rm i} H_0 x }
              ( -{\rm i} ) \int_{0}^{x} {\rm d} s H_1 (s).
 \label{S0+S1}
\end{equation}
We then obtain the oscillation probabilities $P(\nu_{\mu} \rightarrow
\nu_{\rm e})$, $P(\nu_{\mu} \rightarrow \nu_{\mu})$ and $P(\nu_{\mu}
\rightarrow \nu_{\tau})$ in the lowest order approximation as
\begin{eqnarray}
& &
 P(\nu_{\mu} \rightarrow \nu_{\rm e}; L)
=
 4 \sin^2 \frac{\delta m^2_{31} L}{4 E}
 c_{\phi}^2 s_{\phi}^2 s_{\psi}^2
 \left\{
  1 + \frac{a}{\delta m^2_{31}} \cdot 2 (1 - 2 s_{\phi}^2)
 \right\}
 \nonumber \\
&+&
 2 \frac{\delta m^2_{31} L}{2 E} \sin \frac{\delta m^2_{31} L}{2 E}
 c_{\phi}^2 s_{\phi} s_{\psi}
 \left\{
  - \frac{a}{\delta m^2_{31}} s_{\phi} s_{\psi} (1 - 2 s_{\phi}^2)
  +
 \frac{\delta m^2_{21}}{\delta m^2_{31}} s_{\omega}
    (-s_{\phi} s_{\psi} s_{\omega} + c_{\delta} c_{\psi} c_{\omega})
 \right\}
 \nonumber \\
&-&
 4 \frac{\delta m^2_{21} L}{2 E} \sin^2 \frac{\delta m^2_{31} L}{4 E}
 s_{\delta} c_{\phi}^2 s_{\phi} c_{\psi} s_{\psi} c_{\omega}
 s_{\omega},
 \label{mu2e}
\end{eqnarray}
\begin{eqnarray}
 P(\nu_{\mu} \rightarrow \nu_{\mu})
&=&
 1 +
     4 \sin^2 \frac{\delta m_{31}^2 L}{4E}
     c_{\phi}^2 s_{\psi}^2
     \left\{ ( c_{\phi}^2 s_{\psi}^2 - 1 ) +
      \frac{a}{\delta m_{31}^2} \cdot
      2 s_{\phi}^2 (1 - 2 c_{\phi}^2 s_{\psi}^2)
     \right\} \nonumber \\
 &+& 2 \frac{\delta m_{31}^2 L}{2E}
     \sin \frac{\delta m_{31}^2 L}{2E}
     c_{\phi}^2 s_{\psi}^2
     \left\{
        \frac{a}{\delta m_{31}^2}
        s_{\phi}^2 (2 c_{\phi}^2 s_{\psi}^2 - 1)
     \right. \nonumber \\
 &+&
     \left.
      \frac{\delta m_{21}^2}{\delta m_{31}^2}
        (s_{\phi}^2 s_{\psi}^2 s_{\omega}^2
       + c_{\omega}^2 c_{\psi}^2
       - 2 c_{\delta} c_{\psi} c_{\omega} s_{\phi} s_{\psi}
           s_{\omega} )
     \right\}
 \label{mu2mu}
\end{eqnarray}
and
\begin{eqnarray}
 P(\nu_{\mu} \rightarrow \nu_{\tau})
 &=& 4 \sin^2 \frac{\delta m^2_{31} L}{4E}
     c_{\phi}^4 c_{\psi}^2 s_{\psi}^2
     \left(
      1 - \frac{a}{\delta m^2_{31}} \cdot 4 s_{\phi}^2
     \right)
 \nonumber \\
 &+&
  2 \frac{\delta m^2_{31} L}{2E}
  \sin \frac{\delta m^2_{31} L}{2E}
  c_{\phi}^2 c_{\psi} s_{\psi}
  \left[
     \frac{a}{\delta m_{31}^2}
     2 c_{\phi}^2 c_{\psi} s_{\phi}^2 s_{\psi}
  \right. \nonumber \\
 &-&
  \left.
     \frac{\delta m_{21}^2}{\delta m_{31}^2}
     \left\{
      (c_{\omega}^2 - s_{\omega}^2 s_{\phi}^2)
      c_{\psi} s_{\psi}
     +
      c_{\delta} (c_{\psi}^2 - s_{\psi}^2)
      s_{\phi} c_{\omega} s_{\omega}
    \right\}
  \right] \nonumber \\
 &+&
  4 \frac{\delta m^2_{21} L}{2E}
  \sin^2 \frac{\delta m^2_{31} L}{4E}
  s_{\delta} c_{\phi}^2 s_{\phi} c_{\psi} s_{\psi}
  c_{\omega} s_{\omega}.
 \label{mu2tau}
\end{eqnarray}
(Detailed derivation is presented in the \ref{appendix:derivation}).
The transition probabilities for other processes can be written down
explicitly, though we do not present them here.  We chose $\nu_{\mu}$
as initial state allowing for experimental availability.

As is shown in the above transition probabilities, there is matter
effect (proportional to ``$a$'') and we need to distinguish pure
$CP$-violation effect from the fake $CP$ violation due to matter.

\subsection{$T$ Violation}
\label{subsection:disparate-T}
\quad\,
Since $T$ violation is free from the matter effect for the lowest
order,\footnote{For higher order correction due to matter, see
ref.\cite{ArafuneJoe}.}  we first consider how large the $T$-violation
effect can be.  As illustrated in \ref{appendix:derivation}
(the last term of eq.(\ref{alpha2betaapp})), $T$-violation effects are
given by
\begin{equation}
  P(\nu_{\mu} \rightarrow \nu_{\rm e}) -
  P(\nu_{\rm e} \rightarrow \nu_{\mu})
 =
  -8 \frac{\delta m^2_{21} L}{2E}
  \sin^2 \frac{\delta m^2_{31} L}{4E}
  s_{\delta} c_{\phi}^2 s_{\phi} c_{\psi} s_{\psi}
  c_{\omega} s_{\omega}
 \label{T-violation:mu2e}
\end{equation}
and
\begin{equation}
  P(\nu_{\mu} \rightarrow \nu_{\tau}) -
  P(\nu_{\tau} \rightarrow \nu_{\mu})
 =
  8 \frac{\delta m^2_{21} L}{2E}
  \sin^2 \frac{\delta m^2_{31} L}{4E}
  s_{\delta} c_{\phi}^2 s_{\phi} c_{\psi} s_{\psi}
  c_{\omega} s_{\omega},
 \label{T-violation:mu2tau}
\end{equation}
which coincide with eq.(\ref{eq:jf}).  We see the oscillatory part $f$
defined in eq.(\ref{eq:jf}) is given by (see eq.(\ref{eq:fv}))
\begin{equation}
  f \simeq
  2 \Delta_{21} \sin^2 \frac{\Delta_{31}}{2}
  \label{eq:fapp}
\end{equation}
for our approximation.  Here $f \sim {\cal O}(\epsilon \equiv \delta
m_{21} / \delta m_{31}) \ll 1$, since $\Delta_{31} \sim 1$
and $\Delta_{21} \ll 1$ (recall eq.(\ref{eq:Delta-numerical})).

\begin{figure}
 \unitlength=1cm
 \begin{picture}(16,6)
 \unitlength=1mm
 \put(25,80){$f(\Delta_{31},\epsilon = 0.03)$}
 \put(150,5){$\Delta_{31}$}
 \centerline{
 \epsfxsize=13cm
 \epsfbox{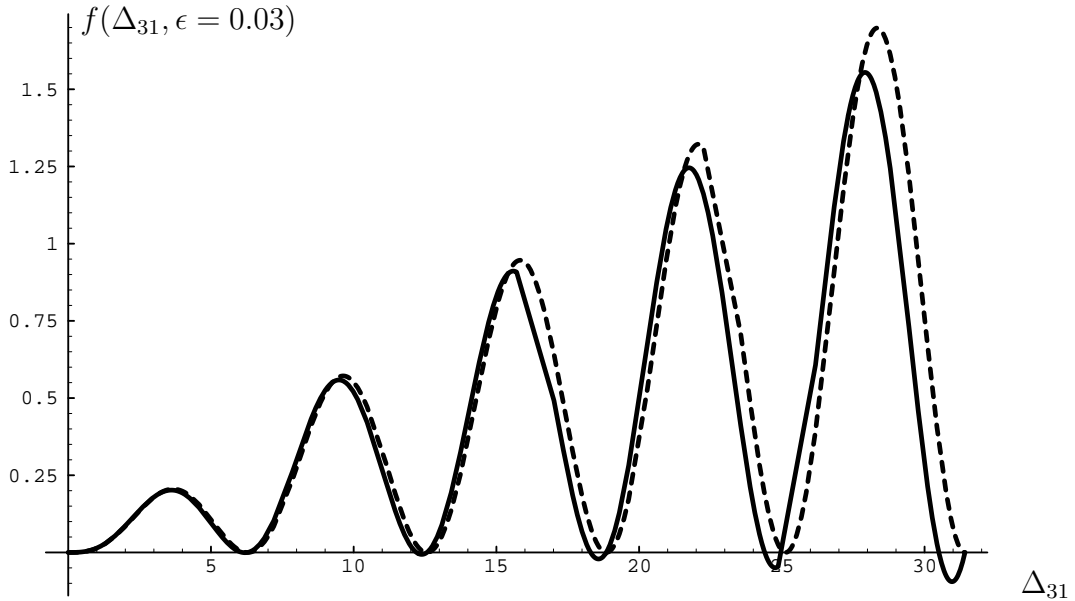}
}
\end{picture}
\caption{Graph of $f(\Delta_{31},\epsilon)$
for $\epsilon = 0.03$.  The solid line and the dashed line represent
the exact expression eq.(\ref{eq:fdef}) and the approximated one
eq.(\ref{eq:fapp}), respectively.  The approximated $f$ has peaks at
$\Delta_{31}=3.67, 9.63, 15.8, \cdots$ irrespectively of $\epsilon$.
}
\label{fig1}
\end{figure}

\begin{table}
 \begin{center}
  \begin{tabular}{c|c|c|c}
   $\Delta_{31}$ & \ $f/\epsilon$\ &
   $\left< f/\epsilon \right>_{10\%}$ &
   $\left< f/\epsilon \right>_{20\%}$\\
   \hline & & &\\
   3.67 & 6.84 & 6.75 & 6.48\\
   9.63 & 19.1 & 17.6 & 14.0\\
   15.8 & 31.5 & 25.7 & 15.6\\
  \vdots&\vdots&\vdots&\vdots
  \end{tabular}
 \end{center}
 \caption{
  The peak values of $f(\Delta_{31},\epsilon) / \epsilon$ and the 
  corresponding averaged values. Here 
  $\left< f/\epsilon \right>_{20\%(10\%)}$ is
  a value of $f(\Delta_{31},\epsilon) / \epsilon$ averaged 
  over the range $0.8 \Delta_{31} \sim 1.2 \Delta_{31}$
  ($0.9 \Delta_{31} \sim 1.1 \Delta_{31}$).
 }
\label{table1}
\end{table}

We show in Fig.\ref{fig1} the graph of $f(\Delta_{31},\delta m_{21} /
\delta m_{31} = 0.03)$.  The approximation eq.(\ref{eq:fapp}) works
very well up to $|\epsilon \Delta_{31}| \sim 1$.  In the following we
will use eq.(\ref{eq:fapp}) instead of eq.(\ref{eq:fdef}).  We see
many peaks of $f(\Delta_{31},\epsilon)$ in Fig.\ref{fig1}.  In
practice, however, we do not see such sharp peaks but observe the
value averaged around there, for $\Delta_{31}$ has a spread due to the
energy spread of neutrino beam ($|\delta\Delta_{31}/\Delta_{31}| =
|\delta E/E|$).  In the following we will assume $|\delta\Delta_{31} /
\Delta_{31}| = |\delta E / E| = 20\%$ \cite{Nishikawa} as a typical
value.

Table \ref{table1} gives values of $f(\Delta_{31},\epsilon)/\epsilon$
at the first several peaks and the averaged values around there.

We see the $T$-violation effect,
\begin{equation}
 \left<
  P(\nu_\alpha \rightarrow \nu_\beta) -
  P(\nu_\beta \rightarrow \nu_\alpha)
 \right>_{20\%}
 =
 4 J \left< f \right>_{20\%}
 =
 J \epsilon \times
 \left\{
  \begin{array}{c}
   25.9 \\
   56.0 \\
   62.4\\
   \vdots
  \end{array}
 \right.
 \label{eq:atbest2}
\end{equation}
for
\begin{equation}
 \Delta_{31} =
 \left\{
  \begin{array}{c}
   3.67 \\
   9.63 \\
   15.8 \\
   \vdots
  \end{array}
 \right.
\end{equation}
at peaks for neutrino beams with 20\% of energy spread.  Note that
the averaged peak values decrease with the spread of neutrino energy.

It depends on $\delta m_{31}^2, L$ and $E$ which peak we can reach.
The first peak $\Delta_{31} = 3.67$ is reached, for example, by
$\delta m_{31}^2 = 10^{-2}$ eV$^2$, $L=250$ km (for
KEK/Super-Kamiokande long baseline experiments) and neutrino energy $E
= 1.73$ GeV.  In this case we see the $T$-violation effect at best of
$|25.9 J \epsilon| \le 2.50 \epsilon$ since we have a bound on $J$ as
eq.(\ref{eq:Jmax}).

\subsection{``$CP$ Violation''}
\label{subsection:disparate-CP}
\quad\,
In practice only $\nu_{\mu}$ and $\bar\nu_{\mu}$ are available by
accelerator.  It is therefore of practical importance to consider pure
$CP$-violation effect through the observation of ``$CP$ violation'',
i.e. difference between $P(\nu_{\alpha} \rightarrow \nu_{\beta})$ and
$P(\bar\nu_{\alpha} \rightarrow \bar\nu_{\beta})$.

Recalling that $P(\bar\nu_{\alpha} \rightarrow \bar\nu_{\beta})$ is
obtained from $P(\nu_{\alpha} \rightarrow \nu_{\beta})$ by the
replacements $a \rightarrow -a$ and $\delta \rightarrow -\delta$, we
have
\begin{eqnarray}
 \Delta P(\nu_{\mu} \rightarrow \nu_{\rm e})
&\equiv&
 P(\nu_{\mu} \rightarrow \nu_{\rm e}; L)
-
 P(\bar\nu_{\mu} \rightarrow \bar\nu_{\rm e}; L)
 \nonumber \\
&=&
 \Delta P_1(\nu_{\mu} \rightarrow \nu_{\rm e}) +
 \Delta P_2(\nu_{\mu} \rightarrow \nu_{\rm e}) +
 \Delta P_3(\nu_{\mu} \rightarrow \nu_{\rm e})
 \label{DeltaPsdef}
\end{eqnarray}
with
\begin{eqnarray}
 \Delta P_1(\nu_{\mu} \rightarrow \nu_{\rm e})
&=& 
 16 \frac{a}{\delta m^2_{31}} \sin^2 \frac{\delta m^2_{31} L}{4 E}
 c_{\phi}^2 s_{\phi}^2 s_{\psi}^2 (1 - 2 s_{\phi}^2),
 \label{DeltaP1def} \\
 \Delta P_2(\nu_{\mu} \rightarrow \nu_{\rm e})
&=& 
 -4 \frac{a L}{2 E} \sin \frac{\delta m^2_{31} L}{2 E}
 c_{\phi}^2 s_{\phi}^2 s_{\psi}^2 (1 - 2 s_{\phi}^2),
 \label{DeltaP2def}
\end{eqnarray}
and
\begin{eqnarray}
 \Delta P_3(\nu_{\mu} \rightarrow \nu_{\rm e})
&=&
 -8 \frac{\delta m^2_{21} L}{2 E}
 \sin^2 \frac{\delta m^2_{31} L}{4 E}
 s_{\delta} c_{\phi}^2 s_{\phi} c_{\psi} s_{\psi} c_{\omega}
 s_{\omega}.
 \label{DeltaP3def}
\end{eqnarray}
Similarly we obtain
\begin{eqnarray}
 & &
 \Delta P(\nu_{\mu} \rightarrow \nu_{\mu})
 \nonumber \\
 &=&
 16 \frac{a}{\delta m^2_{31}}
 \left[
  \sin^2 \frac{\delta m^2_{31} L}{4 E} -
  \frac{1}{4} \frac{\delta m^2_{31} L}{2 E}
   \sin \frac{\delta m^2_{31} L}{2 E}
 \right]
 c_{\phi}^2 s_{\phi}^2 s_{\psi}^2
 \left(
  1 - 2 c_{\phi}^2 s_{\psi}^2
 \right)
 \label{Deltamu2mu}
\end{eqnarray}
and
\begin{eqnarray}
 & &
 \Delta P(\nu_{\mu} \rightarrow \nu_{\tau})
 \nonumber \\
 &=&
 -32 \frac{a}{\delta m^2_{31}}
 \left[
  \sin^2 \frac{\delta m^2_{31} L}{4 E} -
  \frac{1}{4} \frac{\delta m^2_{31} L}{2 E}
   \sin \frac{\delta m^2_{31} L}{2 E}
 \right]
 c_{\phi}^4 s_{\phi}^2 c_{\psi}^2 s_{\psi}^2
 \nonumber \\
 &+&
 8 \frac{\delta m^2_{21} L}{2 E}
 \sin^2 \frac{\delta m^2_{31} L}{4 E}
 s_{\delta} c_{\phi}^2 s_{\phi} c_{\psi} s_{\psi} c_{\omega}
 s_{\omega}.
 \label{Deltamu2tau}
\end{eqnarray}
Here we make some comments.
\begin{enumerate}
\item $P (\nu_{\alpha} \rightarrow \nu_{\beta})$'s and $\Delta P
    (\nu_{\alpha} \rightarrow \nu_{\beta})$'s depend on $L$ and $E$ as
    functions of $L/E$ apart from the matter effect factor $a \, ( = 2
    \sqrt{2} G_{\rm F} n_{\rm e} E )$.
\item At least four experimental data are necessary to determine the
    function $\Delta P (\nu_{\mu} \rightarrow \nu_{\rm e})$, since it
    has four unknown factors: $\delta m^2_{31}, \delta m^2_{21},
    c_{\phi}^2 s_{\phi}^2 s_{\psi}^2 (1-2 s_{\phi}^2)$ and $s_{\delta}
    c_{\phi}^2 s_{\phi} c_{\psi} s_{\psi} c_{\omega} s_{\omega}$.  In
    order to determine all the mixing angles and the $CP$ violating
    phase, we need to observe $P (\nu_{\mu} \rightarrow \nu_{\mu})$
    and $P (\bar\nu_{\mu} \rightarrow \bar\nu_{\mu})$ in addition.
\item $\Delta P (\nu_{\mu} \rightarrow \nu_{\mu})$ is independent of
    $\delta$ and consists only of matter effect term.
\end{enumerate}

``$CP$ violation'', the difference between $P(\nu_{\alpha} \rightarrow
\nu_{\beta})$ and $P(\bar\nu_{\alpha} \rightarrow \bar\nu_{\beta})$,
consists of two effects: pure $CP$-violation effect and matter effect.
We now investigate how we can divide $\Delta P (\nu_{\mu} \rightarrow
\nu_{\rm e})$ into a pure $CP$-violation part and a matter effect part%
\footnote{ It is straightforward to extend the following arguments to
other processes like $\nu_{\mu} \rightarrow \nu_{\tau}$.  We present
the cases of $\nu_{\mu} \rightarrow \nu_{\rm e}$ and $\bar\nu_{\mu}
\rightarrow \bar\nu_{\rm e}$ as examples.  }.  The terms $\Delta P_1
(\nu_{\mu} \rightarrow \nu_{\rm e})$ and $\Delta P_2 (\nu_{\mu}
\rightarrow \nu_{\rm e})$, which are proportional to ``$a$'', are due
to effect of the matter along the path.  The term $\Delta P_3
(\nu_{\mu} \rightarrow \nu_{\rm e})$, which is proportional to
$s_{\delta}$, represents the pure $CP$ violation and indeed coincides
with the $T$ violation, eq.(\ref{T-violation:mu2e}) (We simply call
$\Delta P_i (\nu_{\mu} \rightarrow \nu_{\rm e})$ as $\Delta P_i$
hereafter).  In the following we introduce two methods to separate the
pure $CP$ violating effect $\Delta P_3$ from the matter effect $\Delta
P_1 + \Delta P_2$.

\subsubsection{Observation of Envelope Patterns}
\begin{figure}
 \unitlength=1cm
  \begin{picture}(15,5)
  \unitlength=1mm
  \centerline{
   \epsfysize=5cm
   \epsfbox{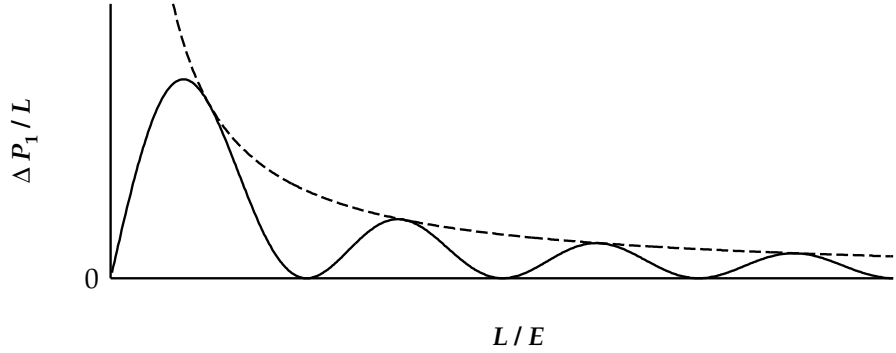}
   } 
\end{picture}
 \begin{flushleft}
     (a) Matter effect term $\Delta P_1(\nu_{\mu} \rightarrow \nu_{\rm
     e})$ divided by $L$ for $c_{\phi}^2 s_{\phi}^2 s_{\psi}^2 (1 - 2
     s_{\phi}^2) > 0$.  The envelope decreases monotonously with $L/E$.
 \end{flushleft}
 \label{x-1sinx}
 \unitlength=1cm
  \begin{picture}(15,5)
  \unitlength=1mm
  \centerline{
   \epsfysize=5cm
   \epsfbox{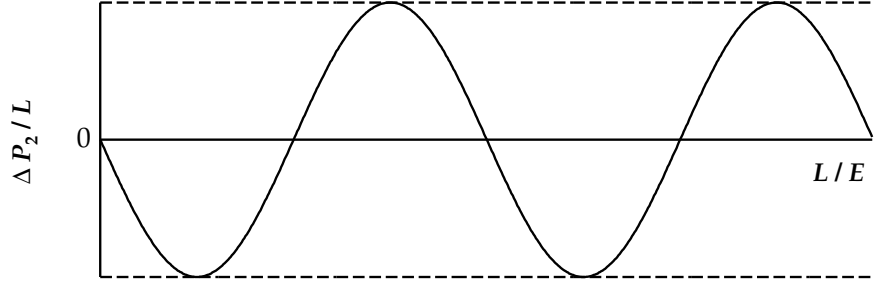}
   }
 \end{picture}
 \begin{flushleft}
     (b) Matter effect term $\Delta P_2(\nu_{\mu} \rightarrow \nu_{\rm
     e})$ divided by $L$ for $c_{\phi}^2 s_{\phi}^2 s_{\psi}^2 (1 - 2
     s_{\phi}^2) > 0$.  The envelope is flat.
 \end{flushleft}
 \label{sinx}
 \unitlength=1cm
  \begin{picture}(15,5)
  \unitlength=1mm
  \centerline{
   \epsfysize=5cm
   \epsfbox{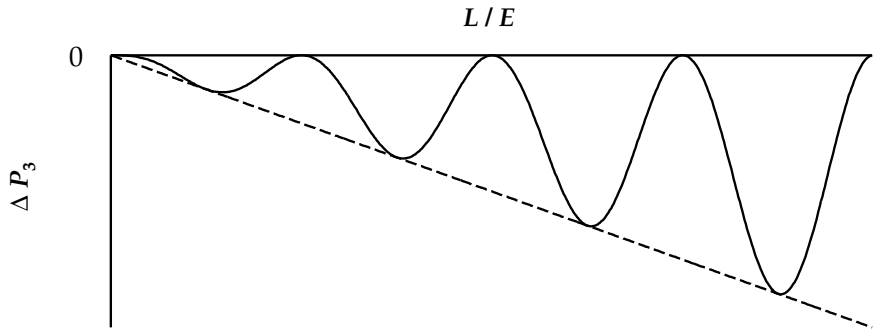}
   }
 \end{picture}
  \begin{flushleft}
   (c) $CP$-violation effect term
       $\Delta P_3(\nu_{\mu} \rightarrow \nu_{\rm e})$ for
       $ s_{\delta} c_{\phi}^2 s_{\phi} c_{\psi} s_{\psi} c_{\omega}
         s_{\omega} > 0$.  The envelope increases linearly with $L/E$.
  \end{flushleft}
 \label{xsinx}
 \caption{The oscillation behaviors of the $\Delta P_1, \Delta P_2$ and 
 $\Delta P_3$.}
 \label{OscBehave}
\end{figure}

\quad\,
One method is to observe the pattern of the envelope of $\Delta P$,
and to separate $\Delta P_3$ from it.  Considering the energy
dependence of $a (\propto E)$, we see that $\Delta P_1 /L$, $\Delta
P_2 /L$ and $\Delta P_3$ depend on a variable $L/E$ alone.  The
dependences of them on the variable $L/E$, however, are different from
each other as seen in Fig. \ref{OscBehave}.  Each of them oscillates
with common zeros at $L/E = 2 \pi n / \delta m^2_{31} (n = 0, 1, 2,
\cdots)$ and has its characteristic envelope.  The envelope of $\Delta
P_1 /L$ decreases monotonously. That of $\Delta P_2 /L$ is flat.  That
of $\Delta P_3$ increases linearly.
It is thus possible to separate these three functions and determine $CP$
violating effect $\Delta P_3$ by measuring the probability $\Delta P$
over wide energy range in the long baseline neutrino oscillation
experiments.  This method has a merit that we can determine the pure
$CP$ violating effect with a single detector.

\begin{figure}
 \unitlength=1cm
 \begin{picture}(13,9)
 \unitlength=1mm
  \centerline{
   \epsfysize=6.5cm
   \epsfbox{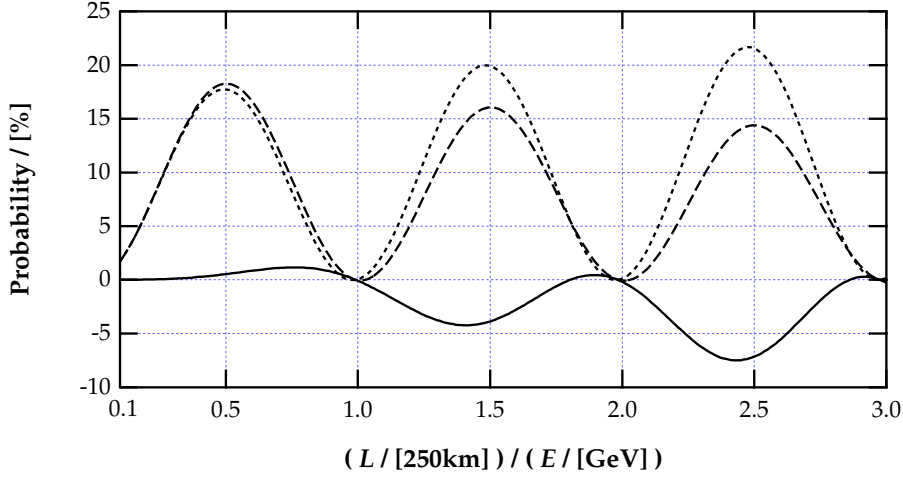}
   }
 \end{picture}
 \begin{center}
     (a) The oscillation probabilities as functions of $L/E$ for
     $\delta = \pi / 2$.
 \end{center}
 \label{mu2emax2}
\vspace{0.3cm}
 \begin{picture}(13,7)
 \unitlength=1mm
  \centerline{
   \epsfysize=6.5cm
   \epsfbox{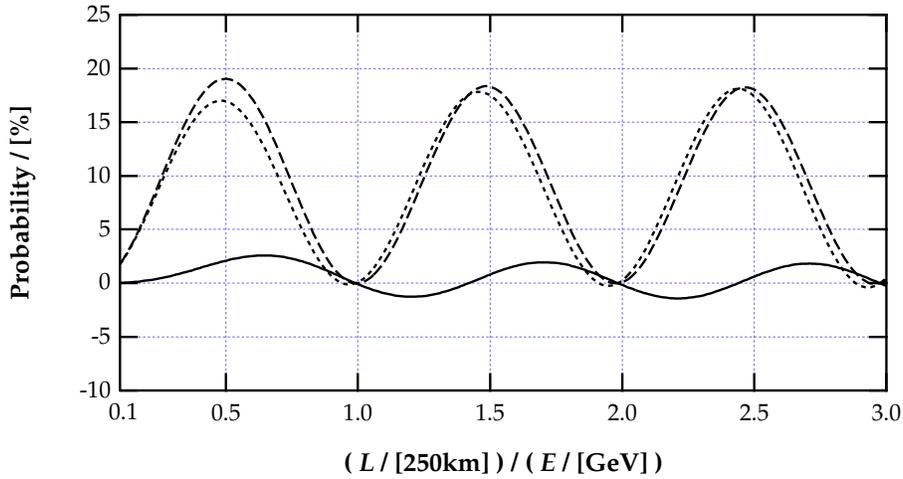}
   }
 \end{picture}
 \label{mu2e02}
 \begin{center}
     (b) The oscillation probabilities as functions of $L/E$ for
     $\delta = 0$.
 \end{center}
 \caption[OscProbCap]
 {The oscillation probabilities for $\delta = \pi / 2$
 (Fig.\ref{OscProb1}(a)) and $\delta = 0$ (Fig.\ref{OscProb1}(b)).
 $P(\nu_{\mu} \rightarrow \nu_{\rm e}), P(\bar\nu_{\mu} \rightarrow
 \bar\nu_{\rm e})$ and $\Delta P(\nu_{\mu} \rightarrow \nu_{\rm e})$
 are given by a broken line, a dotted line and a solid line,
 respectively.  Here $\rho = 2.34 {\rm \,g/cm^3}$ and $L = 250 {\rm
 \,km}$ (the distance between KEK and Super-Kamiokande) are taken.
 Other parameters are fixed at the following values which are
 consistent with the solar and atmospheric neutrino experiments
\cite{FLM}: $\delta m^2_{21} = 10^{-4} {\rm \, eV^2}, \delta m^2_{31}
 = 10^{-2} {\rm \, eV^2}, s_{\psi} = 1/\sqrt{2}, s_{\phi} =
 \sqrt{0.1}$ and $s_{\omega} = 1/2$.}
 \label{OscProb1}
\end{figure}

In Fig.\ref{OscProb1} we give the probabilities $P(\nu_{\mu}
\rightarrow \nu_{\rm e})$ and $P(\bar\nu_{\mu} \rightarrow
\bar\nu_{\rm e})$ for a set of typical parameters which are consistent
with the solar and atmospheric neutrino experiments\cite{FLM}:
$\delta m^2_{21} = 10^{-4} {\rm \, eV^2}, \delta m^2_{31} = 10^{-2}
{\rm \, eV^2}, s_{\psi} = 1/\sqrt{2}, s_{\phi} = \sqrt{0.1}$ and
$s_{\omega} = 1/2$.  We see the effect of pure $CP$ violation in
Fig.\ref{OscProb1}(a), since we find that the curve $\Delta P$ has the
envelope characteristic of $\Delta P_3$.

We comment that the envelope behavior of $\Delta P$ can be understood
rather simply: Since $\Delta P_3$ represents the pure $CP$ violation,
which is same as the $T$-violation effect in the lowest order
of the matter effect,
\begin{equation}
 \Delta P_3 \propto f \simeq 2 \Delta_{21} \sin^2
 \frac{\Delta_{31}}{2}.
 \label{DeltaP3Behavior}
\end{equation}
(See the discussion above the eq.(\ref{eq:fapp})).
This shows $\Delta P_3$ has a linearly increasing envelope
$\Delta_{21} \propto L/E$.  On the other hand, the envelopes of
$\Delta P_1$ and $\Delta P_2$ do not increase with $L/E$ for fixed
$L$, and it makes $\Delta P_3$ dominant in $\Delta P$ for large $L/E$.

Such characters of envelope behaviors enables us to determine whether
$CP$-violation effect is present or not even in case neutrino beam has
energy spread; for neutrino with widely spread energy spectrum, we
observe the average of ``$CP$-violation'' effect which is not zero
if there is pure $CP$-violation effect (see Fig.\ref{OscProb1})%
\footnote{This is also the case for observations of $T$-violation.}.

\subsubsection{Comparison of Experiments with Different $L$'s}
\quad\,
The other method is to separate the pure $CP$ violating effect by
comparison of experiments with two different $L$'s.  Suppose that two
experiments, one with $L = L_1$ and the other $L = L_2$, are
available.  We observe two probabilities $P (\nu_{\mu} \rightarrow
\nu_{\rm e}; E_1, L_1)$ and $P (\nu_{\mu} \rightarrow \nu_{\rm e};
E_2, L_2)$ with $L_1 / E_1 = L_2 / E_2$.  Recalling that $P (\nu_{\mu}
\rightarrow \nu_{\rm e}; L)$ is a function of $L/E$ apart from the
matter effect factor $a (\propto E)$,
we see that the difference
\begin{equation}
 \left\{
  P(\nu_{\mu} \rightarrow \nu_{\rm e}; E_1, L_1) -
  P(\nu_{\mu} \rightarrow \nu_{\rm e}; E_2, L_2)
 \right\}_{L_1/E_1 = L_2/E_2}
 \label{L1-L2}
\end{equation}
is due only to terms proportional to ``$a$''.  We obtain $\Delta P_3$
by subtracting these terms ($\Delta P_1 + \Delta P_2$) from $\Delta
P(\nu_{\mu} \rightarrow \nu_{\rm e})$ as\footnote{Note that the
eq.(\ref{SubMatterEff1}) does not require $P (\bar \nu_{\mu}
\rightarrow \bar \nu_{\rm e}; L_2)$.}
\begin{eqnarray}
& &
 \Delta P_{3} (\nu_{\mu} \rightarrow \nu_{\rm e}; E_1, L_1)
 \nonumber \\
&=&
 \left[
  \Delta P(\nu_{\mu} \rightarrow \nu_{\rm e}; E_1, L_1)
 -
  \frac{2 L_1}{L_2 - L_1}
  \left\{
   P(\nu_{\mu} \rightarrow \nu_{\rm e}; E_2, L_2) -
   P(\nu_{\mu} \rightarrow \nu_{\rm e}; E_1, L_1)
  \right\}
 \right]_{L/E={\rm const.}}
 \label{SubMatterEff1} \\
 &=& \left[ \Delta P(\nu_{\mu} \rightarrow \nu_{\rm e}; E_1, L_1) -
 \frac{L_1}{L_2 - L_1} \left\{ \Delta P(\nu_{\mu} \rightarrow \nu_{\rm
   e}; E_2, L_2) - \Delta P(\nu_{\mu} \rightarrow \nu_{\rm e}; E_1,
 L_1) \right\} \right]_{L/E={\rm const.}}.
 \label{SubMatterEff2}
\end{eqnarray}
This method has a merit that it does not need to observe the envelope
nor many oscillation bumps in the low energy range.

\begin{figure}
 \unitlength=1cm
  \begin{picture}(15,11)
  \unitlength=1mm
 \put(52,103) {$P_{\rm KEK/SK} \equiv P(\nu_{\mu} \rightarrow \nu_{\rm e}; 
 L = 250 {\rm km})$}
 \put(52,97.5){$P_{\rm Minos} \equiv P(\nu_{\mu} \rightarrow \nu_{\rm e}; 
 L = 730 {\rm km})$}
 \put(52,92)  {$P_{\rm KEK/SK} - P_{\rm Minos}$}
 \put(52,86.5){$CP$ violation $\Delta P_3$ for KEK/SK}
  \centerline{
   \epsfysize=11cm
   \epsfbox{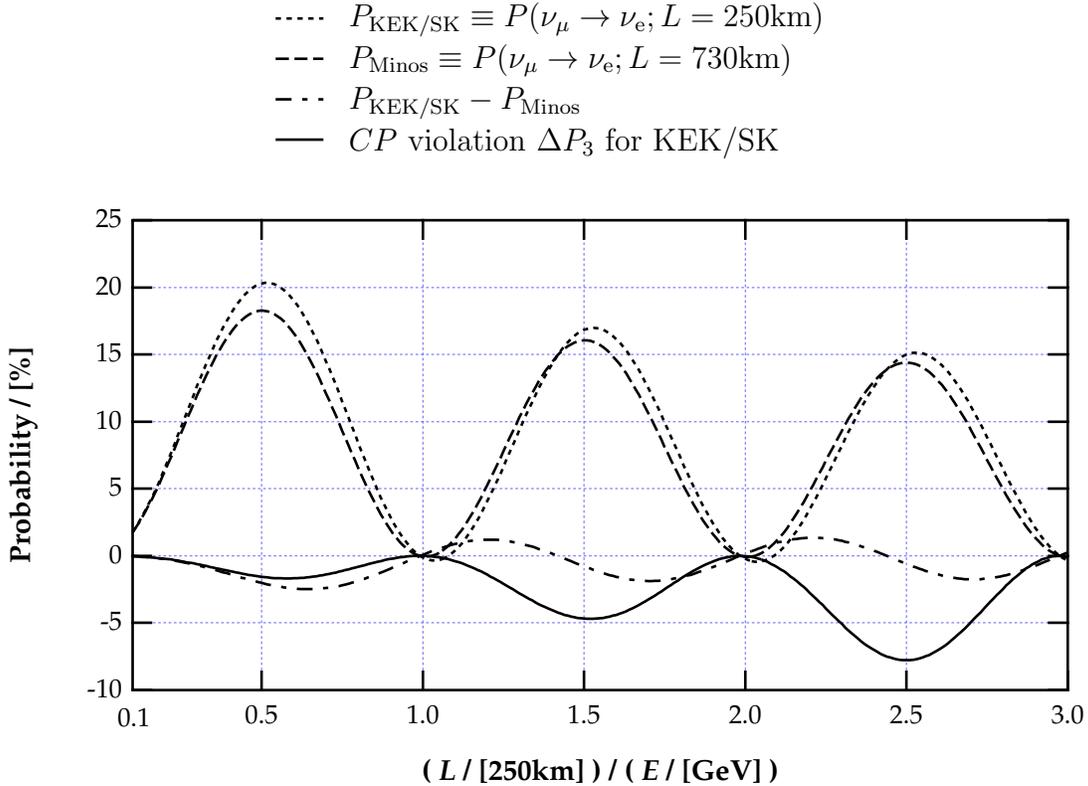}
   } 
\end{picture}
\caption[MinosKekCap]
{The oscillation probabilities $P(\nu_{\mu} \rightarrow \nu_{\rm
e})$'s for KEK/Super-Kamiokande experiments with $L=250 {\rm km}$
(broken line) and those for Minos experiments with $L=730 {\rm km}$
(dotted line).  Masses and mixing angles are the same as in Fig.
\ref{OscProb1}(a).  Their difference, which consists only of matter
effect, is shown by a dot-dashed line.  The pure $CP$ violating effect
in KEK/Super-Kamiokande experiments determined by
eq.(\ref{SubMatterEff1}) is drawn by a solid line.}
 \label{MinosKek}
\end{figure}

In Fig.\ref{MinosKek} we compare $P(\nu_{\mu} \rightarrow \nu_{\rm
e})$ for $L=250 {\rm km}$ (KEK/Super-Kamiokande experiment) with that
for $L=730 {\rm km}$ (Minos experiment) in a case with the same
neutrino masses and mixing angles as those in Fig.\ref{OscProb1}(a).
We see their difference, consisting only of the matter effect, has the
same shape as the solid line in Fig. \ref{OscProb1}(b) up to a
overall constant.  We also show the pure $CP$ violating effect
obtained by the two probabilities with eq.(\ref{SubMatterEff1}).  This
curve has a linearly increasing envelope as seen in Fig.
\ref{OscBehave}(c).

\vspace{1cm}

In this section we have shown that it is possible to determine the
$CP$-violation effect in case $\delta m^2$'s have small
values of ${\cal O}(10^{-4} {\rm eV^2})$ and ${\cal O}(10^{-2} \;{\rm
eV}^2)$, respecting solar neutrino deficit and
atmospheric neutrino anomaly%
\footnote{%
Also some other authors\cite{Tanimoto,MN,BGG} has discussed the
possibility to observe $CP$-violation in the long baseline neutrino
oscillation experiments, but they adopted large $\delta m^2$'s of
${\cal O}(1 \;{\rm eV}^2)$ and ${\cal O}(10^{-2} \;{\rm eV}^2)$,
suggested by 
LSND experiments\cite{LSND} and atmospheric neutrino
observations\cite{Fogli1,FLM,Yasuda}.%
}. Even in this case we may see about 5\% or more $CP$-violation
effect in the near future.

\section{Summary and Discussions}
\label{section:summary}
\quad\,
We have examined the $CP$ and $T$ violation in the neutrino
oscillation and analyzed how large the violation can be, taking into
account the solar neutrino deficit and the atmospheric neutrino
anomaly.

In case of the comparable mass differences with $\delta m^2_{21},
\delta m^2_{31}$ and $\delta m^2_{32}$ in the range $10^{-3}$ to
$10^{-2} {\rm eV^2}$, which is consistent with the analysis of the
atmospheric neutrino anomalies (and maybe with the solar neutrino
deficit), it is found that there is a possibility that the $CP$
violation effect is large enough to be observed by 100 $\sim$ 1000 km
baseline experiments if the $CP$ violating parameter sin$\delta$ is
sufficiently large.

In case that $\delta m^2_{21}$ is much smaller than $\delta m^2_{31}$,
which is favored if we attribute both the solar and atmospheric
neutrino anomalies to the neutrino oscillation, the matter effect by
the earth gives the effective mass equal or greater than the smaller
mass difference $\delta m^2_{21}$ and we cannot ignore the presence of
matter.

We have given very simple formulae for the transition probabilities of
neutrinos in long baseline experiments for this case.  They have taken
into account not only the $CP$-violation effect but also the matter
effect, and are applicable to such interesting parameter regions that
can explain both the atmospheric neutrino anomaly and the solar
neutrino deficit by the neutrino oscillation.

With these simple expressions we have shown that measurement of the
$T$ violation gives the pure $CP$ violating effect.

We have also shown with the aid of these formulae two methods to
distinguish pure $CP$ violation from matter effect.  The dependence of
pure $CP$-violation effect on the energy $E$ and the distance $L$ is
different from that of matter effect: The former depends on $L/E$
alone and has a form $f(L/E)$, while the latter has a form $L \times
g_1(L/E) \equiv E \times g_2(L/E)$.  One method to distinguish is
to observe closely the energy dependence of the difference $
P(\nu_{\mu} \rightarrow \nu_{\rm e}; L) - P(\bar\nu_{\mu} \rightarrow
\bar\nu_{\rm e}; L) $ including the envelope of oscillation bumps.
The other is to compare results from two different distances $L_1$ and
$L_2$ with $L_1/E_1 = L_2/E_2$ and then to subtract the matter effect
by eq.(\ref{SubMatterEff1}) or eq.(\ref{SubMatterEff2}).

Each method has both its merits and demerits.  The first one has a
merit that we need experiments with only a single detector.  A merit
of the second is that we do not need wide range of energy (many bumps)
to survey the neutrino oscillation.

It is desirable to make long baseline neutrino oscillation experiments
with high intensity neutrino flux, and to study $CP$ violation in the
lepton sector experimentally.  Even if the mass differences are very
disparate we may see about 5\% or more $CP$ violation as is seen from
Fig.\ref{OscProb1} and Fig.\ref{MinosKek}.

\section*{Acknowledgments}
\quad\,
We would like to thank J.~Arafune and K.~Hagiwara for their valuable
comments and encouragement.  We are very grateful to M.~Sakuda,
A.~Sakai and M.~Komazawa.

\appendix
\renewcommand{\thesection}{Appendix \Alph{section}}
\setcounter{equation}{0}
\renewcommand{\theequation}{\Alph{section}\arabic{equation}}
\section{Derivation of the Oscillation Probabilities}
\label{appendix:derivation}
\quad\,
Here we present the derivation of eq.(\ref{mu2e}) $\sim$
eq.(\ref{mu2tau}) with use of eq.(\ref{S0+S1}), and show how well this
approximation works.
Let us set $S(x) = S_0(x) + S_1(x)$, defining
\begin{eqnarray}
 S_0 (x) &=& {\rm e}^{-{\rm i} H_0 x},
 \label{S0def} \\
 S_1 (x) &=& {\rm e}^{-{\rm i} H_0 x}
             ({\rm -i}) \int_0^x {\rm d} s H_1 (s).
 \label{S1def}
\end{eqnarray}
We see 
\begin{eqnarray}
 S_0 (x)_{\beta \alpha}
&=&
 \left\{
  U
  {\rm e}^{-{\rm i} \frac{x}{2 E} {\rm diag}(0, 0, \delta m^2_{31})}
  U^{\dagger}
 \right\}_{\beta \alpha}
 \nonumber \\
&=&
 \delta_{\beta \alpha} +
 U_{\beta 3} U^{\ast}_{\alpha 3}
 ( {\rm e}^{ -{\rm i} \frac{\delta m^2_{31} x}{2 E}} - 1)
\end{eqnarray}
and
\begin{eqnarray}
 S_1 (x)_{\beta \alpha}
&=&
 - {\rm i} \int_0^x {\rm d}s \,
 \left[
  {\rm e}^{ -{\rm i} H_0 (x - s) } H_1 {\rm e}^{ -{\rm i} H_0 s }
 \right]_{\beta \alpha}
 \nonumber \\
&=&
  -{\rm i\,} U_{\beta i} U^{\ast}_{\gamma i}
 (H_1)_{\gamma \delta} U_{\delta j} U^{\ast}_{\alpha j}
 \Gamma(x)_{ij},
\end{eqnarray}
where
\begin{eqnarray}
 \Gamma(x)_{ij}
&\equiv&
 \int_0^x {\rm d}s \,
 {\rm e}^{ -{\rm i \,} \frac{\delta m^2_{31}}{2 E}
          \{ (x-s) \delta_{i3} + s \delta_{j3} \} }
 \nonumber \\
&=&
   \delta_{i3} \delta_{j3} \cdot
     x {\rm e}^{ -{\rm i} \frac{\delta m^2_{31} x}{2 E}}
  \nonumber \\
&+&
   \left\{
    (1 - \delta_{i3}) \delta_{j3} +
    \delta_{i3} (1 - \delta_{j3})
   \right\}
   \cdot
     \left( -{\rm i} \frac{\delta m^2_{31}}{2 E} \right)^{-1}
     \left( {\rm e}^{ -{\rm i} \frac{\delta m^2_{31} x}{2 E}} - 1
     \right)
  \nonumber \\
&+&
     (1 - \delta_{i3}) (1 - \delta_{j3}) \cdot x.
\end{eqnarray}
Using
\begin{eqnarray}
 U^{\ast}_{\gamma i} (H_1)_{\gamma \delta} U_{\delta j}
&=&
 \frac{1}{2 E}
 \left\{
  {\rm diag}(0, \delta m^2_{21}, 0) +
  U^{\dagger} {\rm diag}(a, 0, 0) U
 \right\}_{ij}
 \nonumber \\
&=&
 \frac{\delta m^2_{21}}{2 E} \delta_{i2} \delta_{j2} +
 \frac{a}{2 E} U^{\ast}_{1i} U_{1j}
\end{eqnarray}
and
\begin{equation}
 \sum_{k=1}^2
 U^{\ast}_{\alpha k} U_{1 k}
=
 \delta_{\alpha 1} -
 U^{\ast}_{\alpha 3} U_{13},
\end{equation}
we obtain
\begin{equation}
 S(x)_{\beta \alpha}
 =
  \delta_{\beta \alpha} + {\rm i \,} T(x)_{\beta \alpha}
\end{equation}
with
\begin{eqnarray}
 {\rm i \,} T(x)_{\beta \alpha}
&=&
 -2 {\, \rm i\,} {\rm e}^{ -{\rm i} \frac{\delta m^2_{31} x}{4 E} }
 \sin \frac{\delta m^2_{31}}{4 E}
 U_{\beta 3} U^{\ast}_{\alpha 3}
 \left[
   1
  -
   \frac{a}{\delta m^2_{31}}
  \left(
   2 \left| U_{13} \right|^2 - \delta_{\alpha 1} - \delta_{\beta
   1}
  \right)
  -
  {\rm i\,} \frac{a x}{2 E} \left| U_{13} \right|^2
 \right]
 \nonumber \\
&-&
 {\rm i\,} \frac{\delta m^2_{31} x}{2 E}
 \left[
   \frac{\delta m^2_{21}}{\delta m^2_{31}}
   U_{\beta 2} U^{\ast}_{\alpha 2}
  +
 \right. \nonumber \\ & & \left.
   \frac{a}{\delta m^2_{31}}
   \left\{
    \delta_{\alpha 1} \delta_{\beta 1} \left| U_{13} \right|^2
   +
    U_{\beta 3} U^{\ast}_{\alpha 3}
    \left(
    2 \left| U_{13} \right|^2 - \delta_{\alpha 3} -
    \delta_{\beta 3}
    \right )
   \right\}
 \right].
\end{eqnarray}
We then obtain the oscillation probability in the lowest order
approximation as
\begin{eqnarray}
& &
 P (\nu_{\alpha} \rightarrow \nu_{\beta}; L)
=
 \left| S(L)_{\beta \alpha} \right|^2
 \nonumber \\
&=&
 \delta_{\beta \alpha}
 \left[
  1
 -
  4 \left| U_{\alpha 3} \right|^2
  \sin^2 \frac{\delta m^2_{31} L}{4 E}
  \left\{
   1 - 2 \frac{a}{\delta m^2_{31}}
   \left(
   \left| U_{13} \right|^2 - \delta_{\alpha 1}
   \right)
  \right\}
 \right. \nonumber \\ & & \left.
 -
  2 \frac{a L}{2 E} \sin \frac{\delta m^2_{31} L}{2 E}
  \left| U_{\alpha 3} \right|^2
  \left| U_{13} \right|^2
 \right]
 \nonumber \\
&+&
 4 \left| U_{\beta  3} \right|^2
   \left| U_{\alpha 3} \right|^2
 \sin^2 \frac{\delta m^2_{31} L}{4 E}
 \left\{
  1
  -
  4 \frac{a}{\delta m^2_{31}} \left| U_{13} \right|^2
  +
  2 \frac{a}{\delta m^2_{31}} (\delta_{\alpha 1} + \delta_{\beta 1})
 \right\}
 \nonumber \\
&+&
 2 \frac{\delta m^2_{31} L}{2 E} \sin \frac{\delta m^2_{31} L}{2 E}
 \left[
  \frac{\delta m^2_{21}}{\delta m^2_{31}}
  {\, \rm Re\,}
  \left(
   U^{\ast}_{\beta 3} U_{\beta 2}
   U_{\alpha 3} U^{\ast}_{\alpha 2}
  \right)
 \right. \nonumber \\ & & \left.
 +
  \frac{a}{\delta m^2_{31}}
  \left\{
   \delta_{\alpha 1} \delta_{\beta 1} \left| U_{13} \right|^2
   +
   \left| U_{\alpha 3} \right|^2
   \left| U_{\beta 3} \right|^2
   \left(
    2 \left| U_{13} \right|^2
   -
    \delta_{\alpha 1} - \delta_{\beta 1}
   \right)
  \right\}
 \right]
 \nonumber \\
&-&
 4 \frac{\delta m^2_{21} L}{2 E}
 \sin^2 \frac{\delta m^2_{31} L}{4 E}
 {\,\rm Im \,}
 \left(
  U^{\ast}_{\beta 3} U_{\beta 2}
  U_{\alpha 3} U^{\ast}_{\alpha 2}
 \right).
 \label{alpha2betaapp}
\end{eqnarray}
Substituting eq.(\ref{UPar2}) in eq.(\ref{alpha2betaapp}) we finally
obtain eq.(\ref{mu2e}) $\sim$ eq.(\ref{mu2tau}).  Note that all the
terms except the last one in eq.(\ref{alpha2betaapp}) is invariant
under the exchange of $\alpha$ and $\beta$; the last term changes its
sign by this exchange.  It is thus obvious that the last term gives
$T$-violation effect.
\begin{figure}
  \unitlength=1cm
  \begin{picture}(15,8)
  \unitlength=1mm
  \centerline{
   \epsfysize=8cm
   \epsfbox{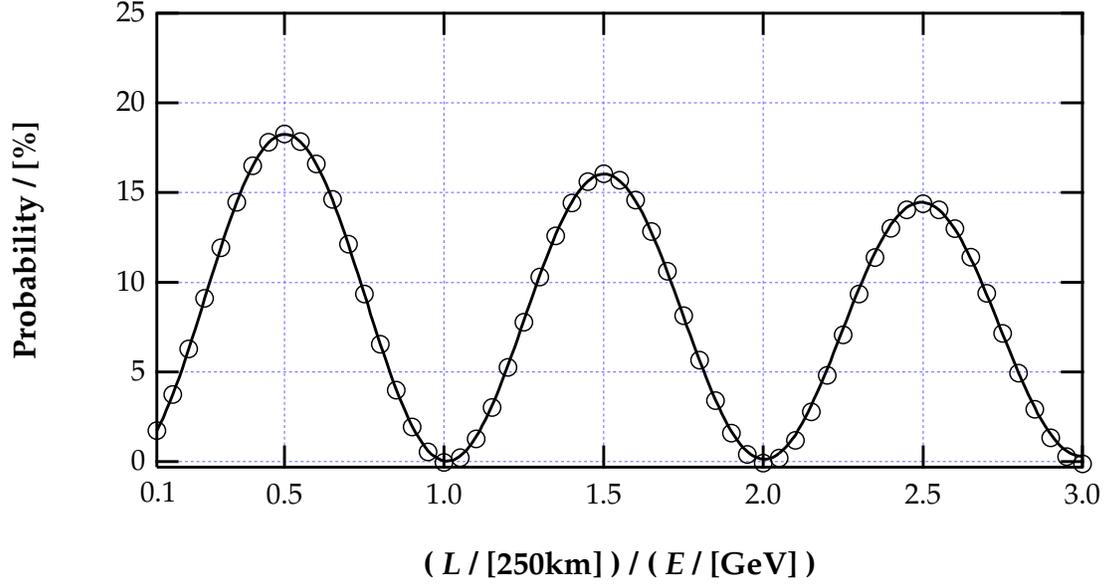}
   } 
\end{picture}
\begin{flushleft}
    {(a) Exact and approximated values of $P (\nu_{\mu} \rightarrow
    \nu_{\rm e})$ for $L = 250$ km assuming constant matter density.}
\end{flushleft}
\begin{picture}(15,8)
  \unitlength=1mm
  \centerline{
   \epsfysize=8cm
   \epsfbox{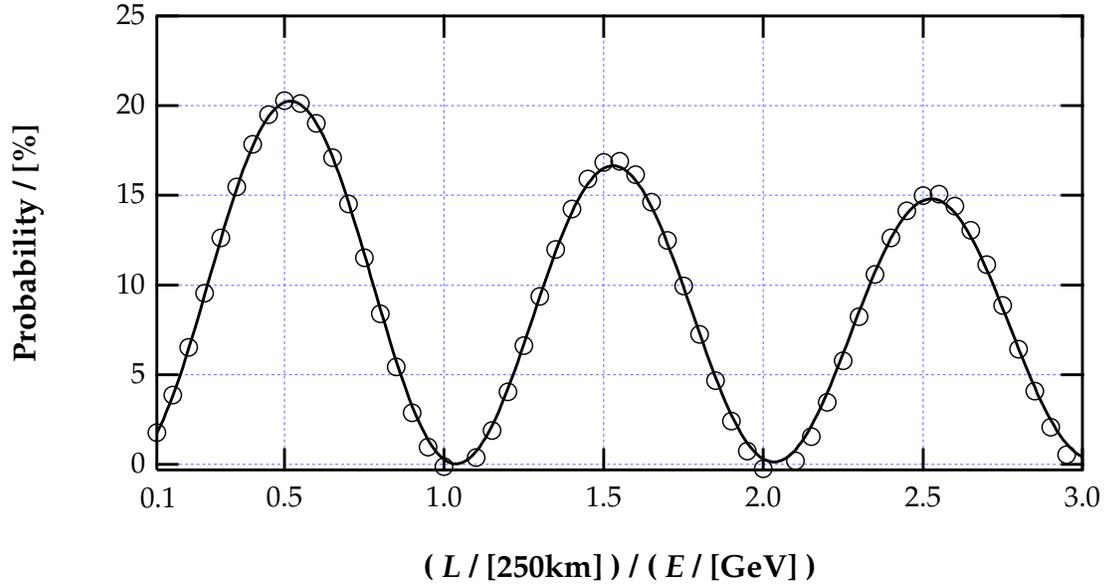}
   } 
\end{picture}
\begin{flushleft}
    {(b) Exact and approximated values of $P (\nu_{\mu} \rightarrow
    \nu_{\rm e})$ for $L = 730$ km assuming constant matter density.}
\end{flushleft}
\caption[ExactApproxCompare]
{Exact and approximated values of $P (\nu_{\mu} \rightarrow \nu_{\rm
e})$ for $L = 250$ km (Fig.\ref{Exact&ApproxCompare}(a)) and those for
$L = 730$ km (Fig.\ref{Exact&ApproxCompare}(b)) assuming constant
matter density.  Exact values and approximated ones are shown by a
solid line and white circles, respectively.  The parameters $s_{\psi},
s_{\phi}, s_{\omega}, \delta$ and $\rho$ are taken the same as in Fig.
\ref{OscProb1}(a).}
 \label{Exact&ApproxCompare}
\end{figure}
\begin{figure}
 \unitlength=1cm
  \begin{picture}(15,8)
  \unitlength=1mm
  \centerline{
   \epsfysize=8cm
   \epsfbox{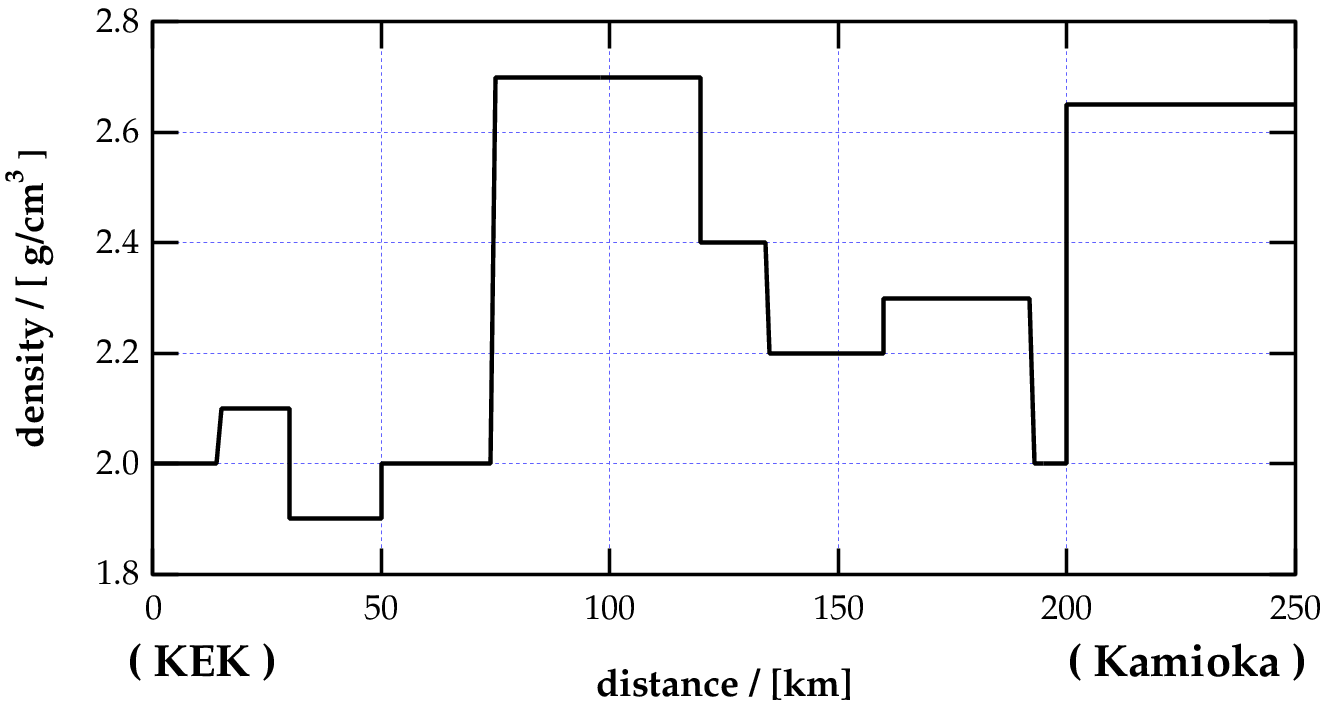}
   } 
\end{picture}
\begin{flushleft}
    {(a) Matter density profile between KEK and
    Super-Kamiokande\cite{Koma}.  Average value of the density is 2.34
    g/cm$^3$.}
\end{flushleft}
\begin{picture}(15,8)
  \unitlength=1mm
  \centerline{
   \epsfysize=8cm
   \epsfbox{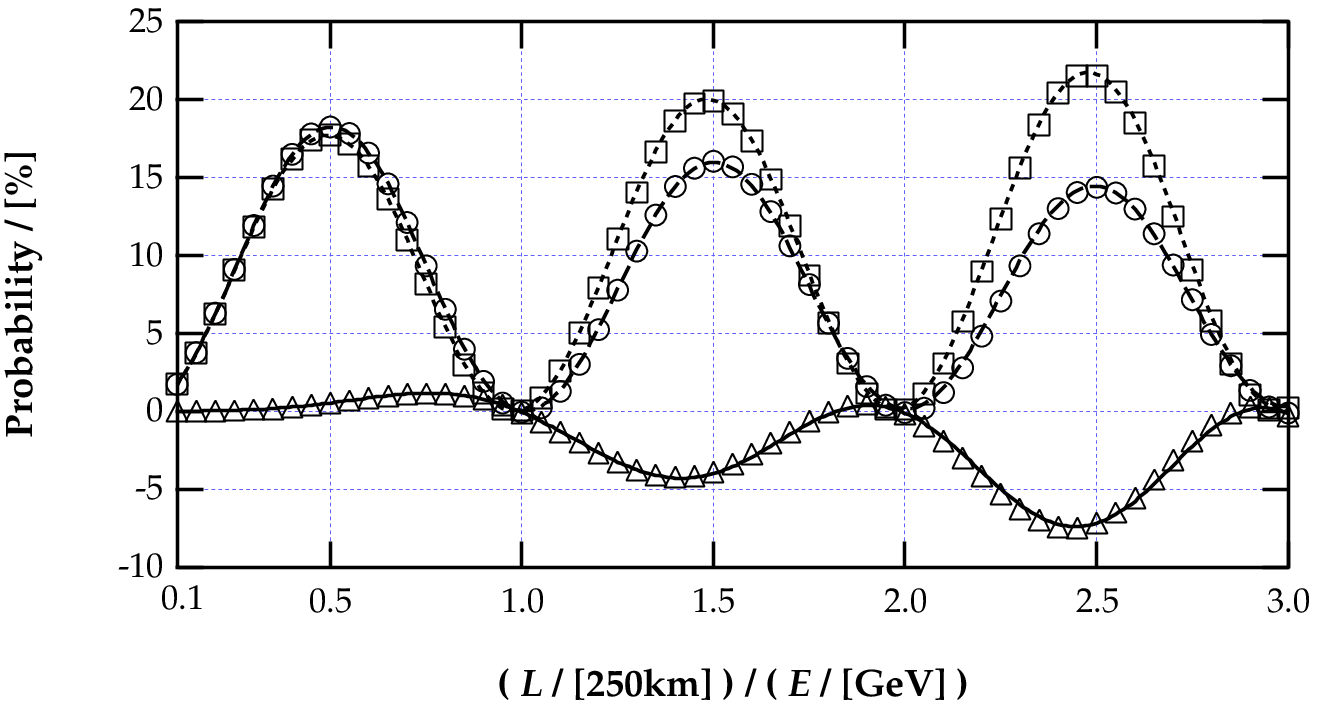}
   } 
\end{picture}
\begin{flushleft}
    {(b) Comparison of values of oscillation probabilities,
    considering and averaging local matter density.  A broken line, a
    dotted line and a solid line are values of $P (\nu_{\mu}
    \rightarrow \nu_{\rm e})$, $P (\bar\nu_{\mu} \rightarrow
    \bar\nu_{\rm e})$ and $\Delta P (\nu_{\mu} \rightarrow \nu_{\rm
    e})$, respectively, taking the density profile shown in (a) into
    account.  Circles, squares and triangles denote the corresponding
    values with constant density approximation (eq.(\ref{mu2e})) with
    averaged matter density, $\rho = 2.34 \; {\rm g/cm^3}$. }
\end{flushleft}
\caption[ExactApproxCompare]
{Effect of matter density variation on $P (\nu_{\mu} \rightarrow
\nu_{\rm e})$, $P (\bar\nu_{\mu} \rightarrow \bar\nu_{\rm e})$ and
$\Delta P (\nu_{\mu} \rightarrow \nu_{\rm e})$.  The parameters
$s_{\omega}, s_{\psi}, s_{\phi}$ and $\delta$ are taken the same as in
Fig.\ref{OscProb1}(a).}
 \label{nonuniformMatter}
\end{figure}

Figure \ref{Exact&ApproxCompare} shows how well this approximation
works for KEK/Super-Kamiokande experiments and also for Minos
experiments with the same masses, mixing angles and $CP$ violating phase
as in Fig.\ref{OscProb1}(a).  Our approximation requires (see
eq.(\ref{AppCond}))
\begin{equation}
 \frac{aL}{2E} =
 0.420 \left( \frac{L}{\rm 730 \,km} \right)
       \left( \frac{\rho}{\rm 3\,g\,cm^{-3}} \right)
 \ll 1
 \label{AppCond1}
\end{equation}
and
\begin{equation}
 \frac{\delta m^2_{21} L}{2E} =
 0.185
 \frac{(\delta m^2_{21} / {\rm 10^{-4} eV^2}) (L/{\rm 730 km})}
      {E / {\rm GeV}}
 \ll 1,
 \label{AppCond2}
\end{equation}
which is marginally satisfied for $L = 730 {\rm km}$.  We see that
even in this case eq.(\ref{alpha2betaapp}) gives good approximation.

We stated that we can use an averaged value $\left< a \right>$ in
place of $a$ in case matter density spatially varies.  We present in
Fig.\ref{nonuniformMatter} the goodness of constant matter density
approximation for KEK/Super-Kamiokande experiments.

\end{document}